# Multicasting with selective delivery: A SafetyNet for vertical handoffs

Henrik Petander[1,2], Eranga Perera[1,3], Aruna Seneviratne[1]

[1]*National ICT Australia, Australian Technology Park, Eveleigh, Australia*

*Email:henrik.petander@nicta.com.au, aruna.seneviratne@nicta.com.au, eranga.perera@nicta.com.au*

[2]*Department of Computer Science, Helsinki University of Technology, 02015 Espoo, Finland*

[3]*Department of Computer Science, University of New South Wales, Kensington, Australia*

In future mobility support will require handling roaming in heterogeneous access networks. In order to enable seamless roaming it is necessary to minimize the impact of the vertical handoffs. Localized mobility management schemes such as FMIPv6 and HMIPv6 do not provide sufficient handoff performance, since they have been designed for horizontal handoffs. In this paper, we propose the SafetyNet protocol, which allows a Mobile Node to perform seamless vertical handoffs. Further, we propose a handoff timing algorithm which allows a Mobile Node to delay or even completely avoid upward vertical handoffs. We implement the SafetyNet protocol and compare its performance with the Fast Handovers for Mobile IPv6 protocol in our wireless test bed and analyze the results. The experimental results indicate that the proposed SafetyNet protocol can provide an improvement of up to 95% for TCP performance in vertical handoffs, when compared with FMIPv6 and an improvement of 64% over FMIPv6 with bicasting. We use numerical analysis of the protocol to show that its signaling and data transmission overhead is comparable to Fast Mobile IPv6 and significantly smaller than that of FMIPv6 with bicasting.

## Key words:

Vertical handoffs, Fast Handovers for Mobile IPv6, bicasting

## 1 Introduction

With the proliferation of new wireless access network technologies, there will be an increasing demand for the capability to roam between heterogeneous networks. Mobile devices, such as PDAs and smartphones are beginning to incorporate multiple network interfaces to enable this roaming. Wireless Local Area Networks (WLANs), e.g. IEEE 802.11 networks, and Wireless Wide Area Networks (WWANs), e.g. WiMax form a wireless overlay as defined by Katz et al.[1]. A mobile user moving within this overlay can take advantage of the complementing characteristics of high speed and low cost of WLAN networks and the larger coverage of WWAN networks.

Roaming in heterogeneous networks results in vertical handoffs between access networks. Such handoffs have a potentially large impact on on-going connections as shown in [2]. In the case of roaming between heterogeneous networks of the same operator or cooperating operators, a localized mobility management protocol can be used to reduce the impact of handoffs. However, the current standardized approaches towards localized mobility management, Fast Handovers for Mobile IPv6[3] and Hierarchical Mobile IPv6[4], do not provide sufficient handoff performance in vertical handoffs, since they have been designed for horizontal handoffs.

Due to the different capabilities of the networks, *upward vertical handoffs* from high speed WLANs to low speed WWANs need to be handled differently from *downward vertical handoffs* from WWANs to WLANs. Downward vertical handoffs are often performed opportunistically, i.e. a Mobile Node performs a handoff to a new network in spite of the current network still being available. Therefore, the handoff timing and duration are often not critical to network application performance. However, in upward handoffs the Mobile Node is typically using the best available network (WLAN) in terms of speed and price and needs to perform a handoff to a WWAN due to moving outside the coverage area of the WLAN. In this case, the handoff latency and timing become crucial. An early handoff would result in unnecessary costs and lower performance from the use of the WWAN network, and a late handoff would result in application performance suffering from packet loss.

This paper has three main contributions. Firstly, we introduce a localized mobility management protocol, SafetyNet which allows a Mobile Node to perform seamless vertical handoffs. This is possible, since the basic SafetyNet mechanism allows the Mobile Node to recover any lost packets when it connects to the new network with a minimal delay, without creating additional over the air traffic, when compared with traditional bicasting or multicasting approaches to seamless handoffs.

Secondly, we propose a vertical handoff timing algorithm for delaying upward vertical handoffs from a low cost network to a higher cost network without degrading the performance of ongoing communications.

Thirdly, we perform an experimental comparison of our implementation of the SafetyNet protocol with the standard Fast Handovers for Mobile IPv6 protocol[3] and the Fast Handovers for Mobile IPv6 with Bicasting protocol[5] in a wireless test bed. The experimental results show that our approach improves the handoff performance significantly over the other protocols, with the performance improvement for TCP ranging between 64% and 95% in the presence of link saturation. We analyze numerically the SafetyNet protocol and compare its overheads with that of Fast Handovers for Mobile IPv6 with and Fast Handovers for Mobile IPv6 with bicasting. The analysis shows that the use of the SafetyNet protocol can reduce the total over the air overhead of a handoff significantly.

# 2 Background and related work

In this section, we discuss related work on vertical handoff performance improvements done on the IP layer. We present an overview of the Fast Handovers for Mobile IPv6 protocol[3] (FMIPv6), since our proposal builds on it and since we use it and FMIPv6 with bicasting[5] in evaluating the performance of our proposal. Additionally, we present previous proposals on vertical handoff timing algorithms.

## 2.1 Localized Mobility Management Protocols

There have been many proposals to improve the performance of IP layer mobility management, such as Hierarchical Mobile IPv6[4], Cellular IP[6]. These mechanisms are focused towards reducing the handoff latency by localizing the handoffs within a network domain. A more complete approach is provided by FMIPv6 which mitigates the impact of handoff by localized forwarding and context transfer between the Access Routers and proactive handoffs. Bicasting or multicasting with simultaneous bindings[5] can be used with FMIPv6 to improve the handoff performance.

FMIPv6 allows a Mobile Node to perform a predictive handoff from its *previous Access Router (pAR*) to a *new Access Router* (*nAR*), if it can anticipate the handoff event before disconnecting from the pAR. In the case where the Mobile Node loses connectivity with the pAR before performing a predictive handoff, the Mobile Node will perform a reactive handoff after connecting to the nAR. In both predictive and reactive modes, the Mobile Node establishes forwarding from its *previous Care-of Address* (*pCoA*) on the link of the pAR to its *new Care-of Address* (*nCOA*) on the link of the nAR. As a part of the handoff, the pAR and the nAR exchange state information for the Mobile Node, such as Quality-of Service (QoS) state, network access service state and security associations. This state transfer mitigates the need for the Mobile Node and the nAR to establish the state after the Mobile Node has connected to a nAR. The state transfer together with the localized forwarding scheme reduces the handoff latency even in the case of reactive handoffs.

A predictive handoff in FMIPv6 consists of the following steps:

1. Discovery of routers connected to the Access Points in the vicinity of the Access Point the Mobile Node is connected to using proxy router solicitation and advertisement messages.
2. Initiation of the handoff via the pAR. The Mobile Node proposes a CoA at link of the nAR in *Fast Binding Update* (FBU), which pAR relays to nAR in a *Handoff Initiate* message. The nAR informs the status of the update with a *Handoff Acknowledge* message. If the handoff is accepted by the nAR, the pAR starts tunneling packets destined to the previous CoA to the new CoA. During the handoff, the nAR receives the tunneled packets and buffers them.
3. Attachment to the link of the nAR. After performing a link layer handoff to the nAR, the Mobile Node sends a *Fast Neighbor Advertisement* to the nAR. Upon receiving the message, the nAR delivers any buffered packets to the Mobile Node.

The tunneling of packets from pCoA to nCoA at the nAR during the handoff prevents the Mobile Node receiving them until the attachment at the nAR. However, the pAR can use bicasting of packets[5] to deliver packets both directly to pCoA and to nCoA during the handoff period. In the proposal, a timer is used for ending the bicasting.

In reactive mode, the Mobile Node forms a new CoA and sends a Fast Binding Update upon connecting to the nAR. The pAR and the nAR then perform context transfer and the pAR starts tunneling packets from the previous CoA to the new CoA.

### 2.2 Vertical Handoff Timing

Several different approaches to vertical handoff timing have been proposed in the literature. Guo et al. propose the use of fuzzy logic and neural networks to optimize the timing to use multiple rules for the handoff decision, including number of users in the candidate networks[7]. In their paper[8], Vidales et al propose the use of concepts from autonomous systems design, including finite-state transducers for improving handoff decisions. Our proposed handoff timing algorithm uses packet loss and application state to delay or in the best case completely avoid an upward vertical handoff without degrading application performance.

# 3 The SafetyNet approach

In this section, we give an overview of the SafetyNet protocol which enables seamless vertical handoffs using a localized mobility management scheme. We design a handoff timing algorithm which exploits the SafetyNet protocol to allow a Mobile Node to maximize its use of low cost networks by delaying a vertical handoff without degrading application performance.

SafetyNet allows a Mobile Node to perform a lossless vertical handoff by introducing a localized mobility management scheme for vertical handoffs. SafetyNet improves our previously proposed Fast Handovers for Mobile IPv6 Bicasting with Selective Delivery [9] (FMIPv6-BSD), a protocol for seamless horizontal handoffs, for vertical handoffs. In SafetyNet, the current Access Router (pAR) starts multicasting packets to candidate Access Router(s) as well as to the Mobile Node at the initialization of the handoff to ensure that any packets lost during the handoff can be recovered. Packets lost during the handoff are delivered to the Mobile Node at the finalization of the handoff from the buffer of the new Access Router. The selective delivery mechanism employed in the protocol ensures that only the lost packets are delivered from the buffer, as opposed to the entire contents of the buffer. This decreases the data transmission overhead of the protocol significantly.

Using a vertical handoff timing algorithm based on the signal strength, the Mobile Node would finalize the handoff immediately after the initialization of the handoff. As opposed to this, in the SafetyNet handoff timing algorithm, the Mobile Node would keep using the pAR until it 1) would arrive at a new nAR of the preferred lower cost network type, allowing it to avoid a vertical handoff and instead perform a horizontal handoff or until it 2) would lose an amount of packets deemed intolerable to the application which would require a vertical handoff. When either of these conditions is met, the Mobile Node would attach to the selected nAR and finalize the handoff. The use of the SafetyNet handoff algorithm for upward vertical handoffs together with the SafetyNet protocol would allow the Mobile Node to maximize its use of the lower cost network without degrading the performance of on-going connections.

Using Figure 1 we give an overview of the operation of the SafetyNet timing algorithm. The handoff timing algorithm would work in the following manner for a Mobile Node depicted in the figure. The Mobile Node would initialize the handoff towards both the WLAN 2 and WWAN Access Routers. It would then delay the finalization of the handoff until either of the two conditions described in the previous paragraph is met. In the case of the Mobile Node moving on Path A, the delaying of the finalization of the handoff would allow it to perform a horizontal handoff to the nAR of WLAN2. In the case of Path B, the Mobile Node would eventually perform a handoff to the nAR of the WWAN. The timing of the handoff finalization is described in the SafetyNet handoff timing algorithm. The delaying of the handoff finalization would incur some packet loss. However, the use of the SafetyNet protocol would allow the Mobile Node to recover any packets lost during the handoff.

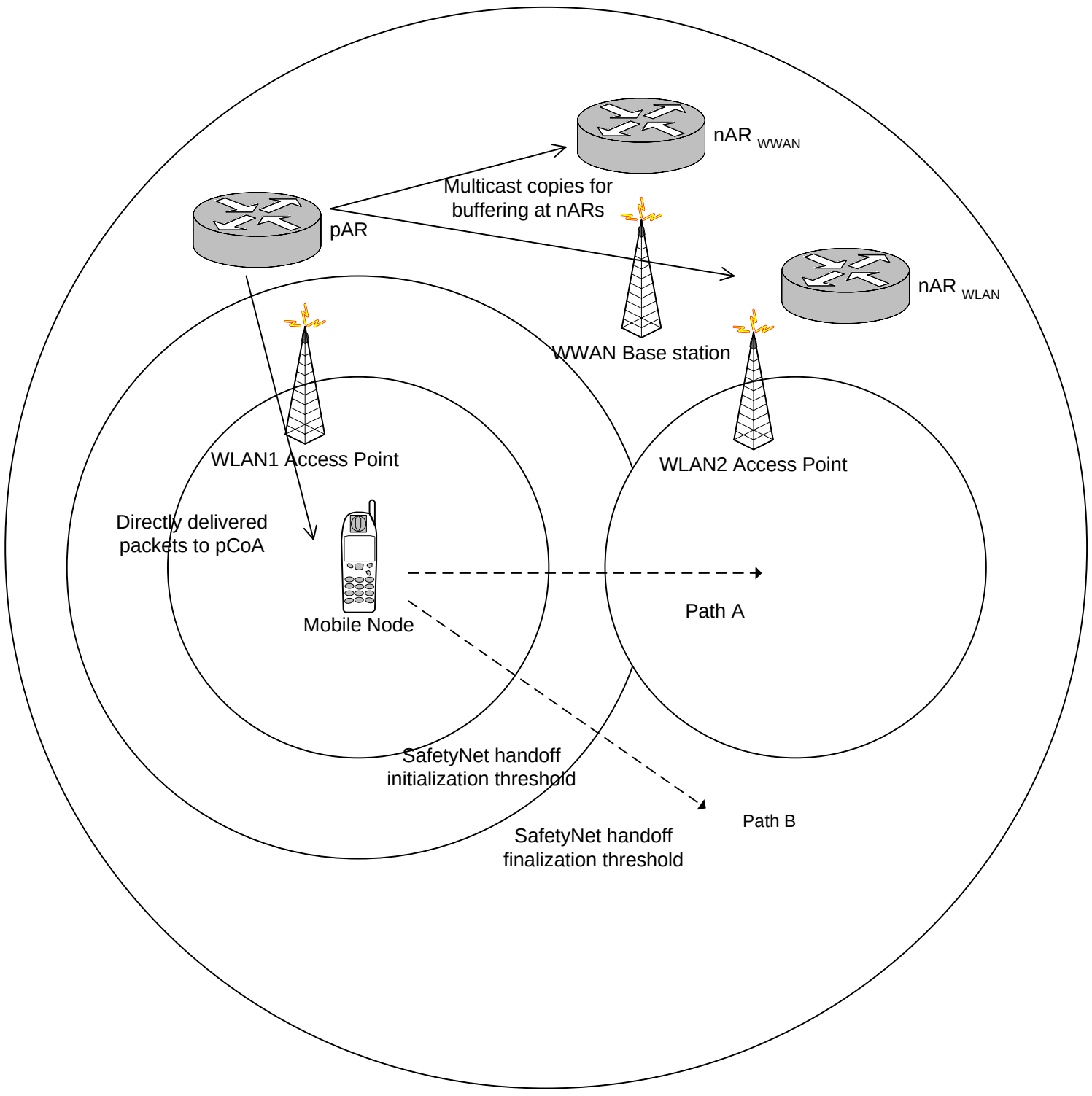

**Figure 1. The SafetyNet mechanism**

## 3.1 A Detailed Description of the SafetyNet protocol

In the SafetyNet protocol, a Mobile Node would initiate a vertical handoff due to degradation of signal strength or the availability of a better network. It would then perform proxy router discovery via the current Access Router (pAR) in order to learn the information, including the IP addresses for detected Access Points or Base Stations.

The Mobile Node would not necessarily be able to detect all the relevant Access Routers, as depicted in Figure 1. The pAR would inform the Mobile Node of the best candidate nARs based on location tracking information of the Mobile Node derived from signal strength together with triangulation, as proposed in [10]. Once this information was obtained by the Mobile Node, it would initiate a handoff with the chosen nARs.

After the Proxy Router Discovery, the Mobile Node would send a Fast Binding Update message to the pAR indicating the selected nAR(s). The pAR would set up multicasting of packets destined to the pCoA of the Mobile Node to the selected nARs by sending a Handoff Initiate message to each nAR. After this initialization of the handoff, the pAR would start to deliver copies of every packet destined to the pCoA of the Mobile Node to the nARs, which would then buffer the packets. Additionally, the pAR would deliver a copy of each packet directly to the pCoA of the Mobile Node. The pAR would mark both the directly delivered and multicast versions of the packets with a counter value incremented for every packet. The counter value would enable the Mobile Node to determine any packets that it did not receive during the handoff and detect any duplicates. This functionality improves the FMIPv6 BSD protocol by enabling a Mobile Node to perform a handoff towards multiple Access Routers.

The Mobile Node would perform the final selection of the Access Router during the handoff based on the handoff timing algorithm presented in the next subsection. Once the nAR was selected, the Mobile Node would send a Fast Neighbor Advertisement to the nAR. The message would contain a list of packets that the Mobile Node did not receive. The nAR would then deliver these missed packets from its buffer and send a *Stop Bicasting* message to the pAR requesting it to stop multicasting packets and send them only to the nAR. Additionally, the Mobile Node would send a Stop Bicasting message to the pAR directly via the link of the pAR. The direct sending of the Stop Bicast message improves the performance over the FMIPv6 BSD protocol by minimizing the amount of duplicate packets sent by pAR after the end of the handoff. The operation of the SafetyNet protocol is illustrated in Figure 2.

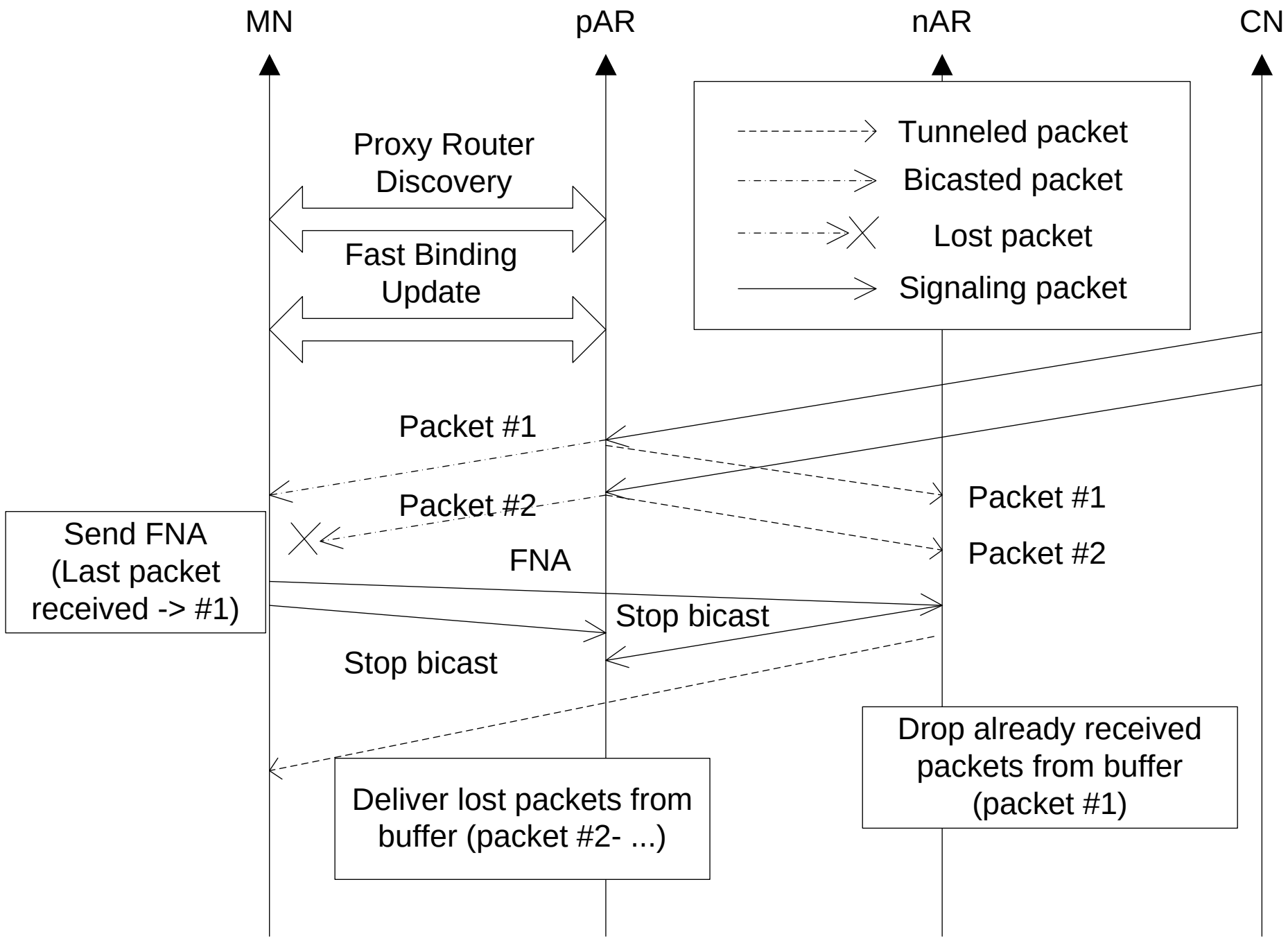

**Figure 2. Operation of the SafetyNet protocol.**

## 3.2 The SafetyNet handoff timing algorithm

The algorithm given below takes into consideration the requirements of the applications being executed in the Mobile Node in order to correctly time an upward vertical handoff. We assume that when the signal strength for the current Access Point or Base Station is sufficient, no significant packet loss will be visible to applications. For example in the case of IEEE 802.11b, we assume that a packet may be occasionally lost due to bit errors, but link layer resending would deliver it with a very high probability and no significant packet loss would be visible to the network layer. When a Mobile Node moves further away from the Access Point, it will eventually start losing packets with an increasing rate, due to the degradation in the received signal strength. This increased error rate at the link layer would at some point exceed the recovery capabilities of the resending mechanism, and would be visible to the network layer. The tolerability of the resulting packet errors would depend on the requirements of the application being executed. When the errors exceeded a tolerance value, the Mobile Node would perform an upward vertical handoff.

We consider a TCP data transfer as a packet loss intolerant example application to demonstrate the operation of the algorithm. With TCP, the loss of a packet starts affecting the performance of the transfer, if the sender does not receive an acknowledgment for the segment during the transfer window. The transfer window starts from the previous acknowledgment. We can derive a tolerable time ($T_t$) that we can delay the finalization of the handoff by calculating as follows:

$$T_t = W_r / B_{pAR} - t_{MN\text{-}CN} - t_d$$

where $W_r$ denotes the remaining transfer window, $B_{pAR}$ the bottleneck bandwidth for the TCP stream before the handoff, $t_{MN\text{-}CN}$ the network latency between the Correspondent Node and the Mobile Node after the handoff via the nAR and $t_d$ the delivery time of packets from buffer of nAR.

The delivery time of packets from the buffer of the nAR consists of the two way propagation delay between the Mobile Node and the nAR and the transmission delay (size of the packets that need to be delivered divided by the bandwidth of the link). TCP knows the remaining transfer window and TCP knows the current round trip time between the Correspondent Node and the Mobile Node which can be used to estimate the $t_{MN\text{-}CN}$. The Mobile Node can estimate the delivery time of packets from the buffer of nAR based on the number and size of any lost packets and the bandwidth of the network technology with the assumption that there is no congestion at the link of the nAR.

# 4 Experimental Evaluation of the SafetyNet Protocol

## 4.1 Implementation and test bed

We developed an experimental prototype of the SafetyNet protocol for the Linux operating system. The prototype extends the fmipv6.org FMIPv6 implementation[11]. The prototype consists of four major parts: signaling in the Mobile Node and the Access Routers, multicasting from the pAR, keeping track of received packets in the Mobile Node and flushing redundant from the buffer at the nAR.

In order to verify the performance of the SafetyNet protocol, we built a wireless test bed. The test bed shown in Figure 3 consists of two Access Routers and a Correspondent Node connected together using a 100Mbits/s Ethernet switch. Both Access Routers serve a WLAN network. In our test bed, in order to emulate vertical handoffs using the available networks, we equip the Mobile Node with two WLAN interfaces. As in a real vertical handoff, the wireless link of the nAR acts as a bottleneck for the traffic. In a WLAN-WWAN handoff, the bottleneck would be more significant due to the asymmetric speeds of the access technologies. However, the effect of the bottleneck seen in our results indicates how the protocols would perform in a WLAN-WWAN handoff.

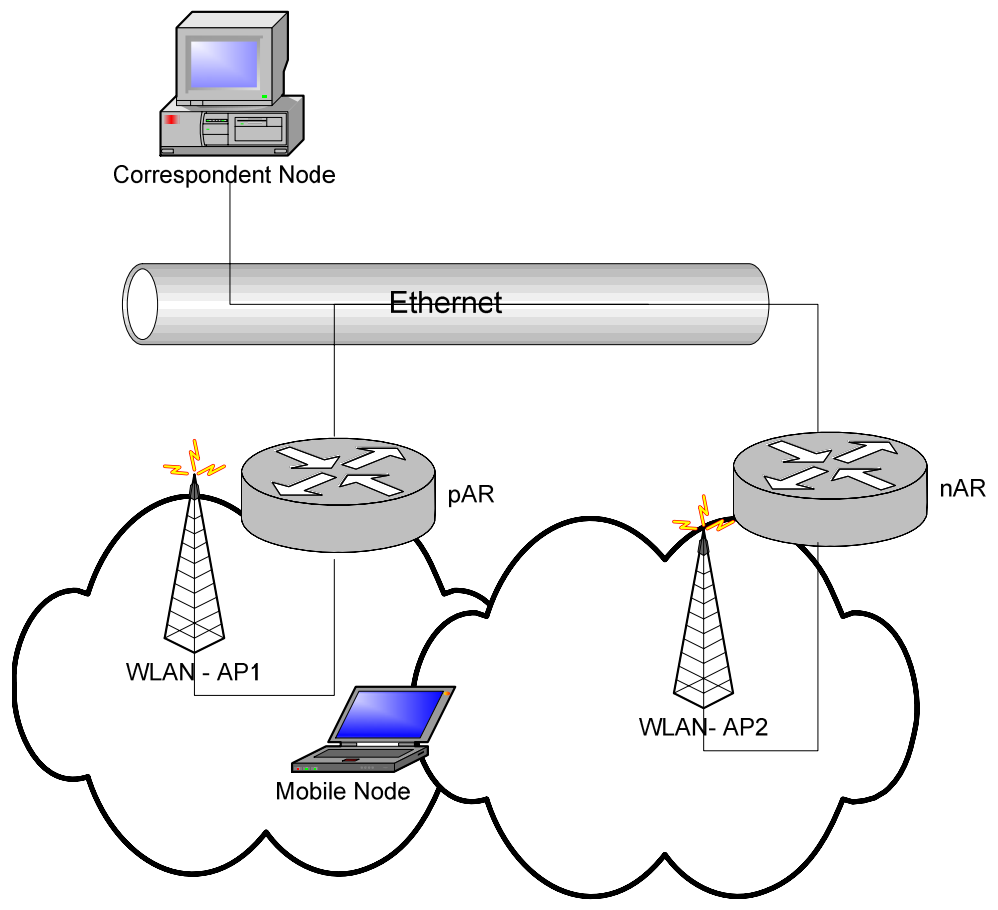

**Figure 3 SafetyNet test bed.**

The two Access Routers were Linux PCs running the SafetyNet and FMIPv6 software. The routers acted as Access Points in managed mode and were set to use the same ESSID, but different channels (1 and 11) to minimize interference. Mobile Node was a Linux laptop running the SafetyNet and FMIPv6 software. The Mobile Node was equipped with two 802.11b interfaces, an integrated Intel IPW2100 card and a Prism 2.5 based PCMCIA card to emulate vertical handoffs. An external antenna was used with the PCMCIA card to minimize the interference from the two co-located cards. Standard Linux drivers without modifications were used for the WLAN cards and the 802.11b link layer attachment latency was on the average 200ms.

We used constant bit rate UDP stream of 100kbits/s with a payload of 100 bytes and a TCP flow as traffic in our experiments. In both cases, the Correspondent Node acted as the source of the traffic and the Mobile Node as the sink. In all experiments, we used a 10s measurement period with the Mobile Node performing a handoff approximately at t=4s. The UDP performance was measured as received data rate at the Mobile Node and for TCP we measured the sequence number progression at the Mobile Node. The TCP sequence number progression was used since it accurately depicts the progress of the data transfer.

## 4.2 Performance Comparison with FMIPv6 and FMIPv6 with Bicasting

In this subsection, we compare the performance of the SafetyNet protocol with FMIPv6 and FMIPv6 with Bicasting. We first use a lower data rate comparison with UDP constant bit rate traffic. The results for UDP handoff performance are presented in Figure 4. FMIPv6 starts delivering packets to buffer of the nAR after the handoff initialization. Thus, after the initialization of the handoff, the Mobile Node cannot receive any packets until it attaches to the nAR, which delivers the packets for the handoff period from its buffer. This can be seen as the drop in received data rate approximately at t=3.8s and the sharp increase approximately at t=4s.

FMIPv6 with bicasting does not suffer from the drop in the packet rate due to the Mobile Node receiving packets directly from the pAR. With the low data rate used for UDP, the delivery of the duplicate packets from the buffer does not affect the performance. However, the high number of duplicate packets could cause performance problems to applications without application level detection of duplicates.

The early stopping of the bicasting in the SafetyNet protocol causes a decrease in the data rate of the UDP traffic (approximately at t=4.5s), which is later (approximately at t=4.7) offset by delivery of traffic from the buffer of nAR.

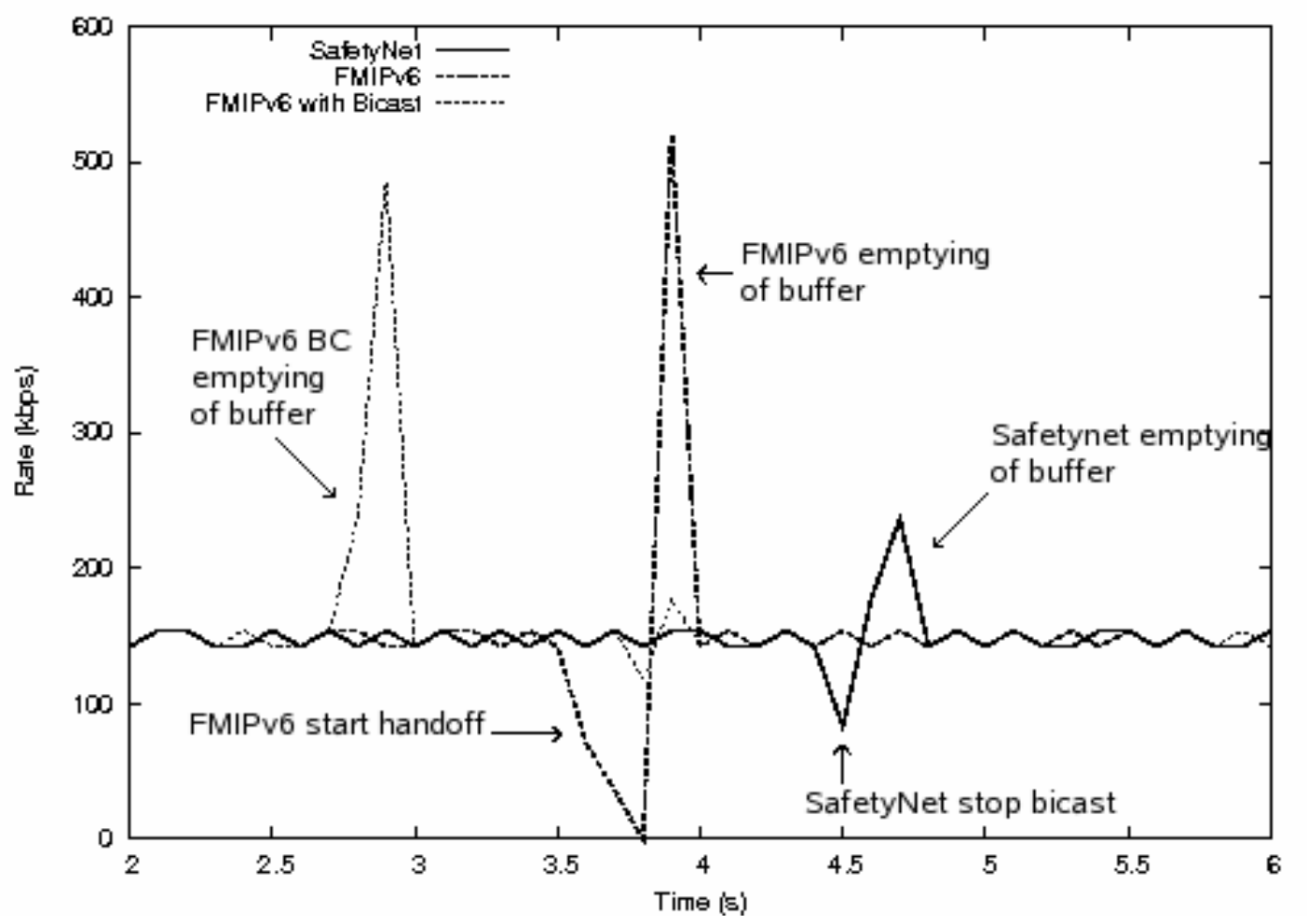

**Figure 4. UDP handoff performance comparison.**

An overview of TCP performance during a handoff for SafetyNet, FMIPv6 and FMIPv6 with bicasting is presented in Figure 5, which shows every 20th TCP sequence number for improving the readability. The three TCP flows progress with the same rate before and after the handoff, with the only difference coming from the handoff performance. We measured the effect of the handoff on TCP by comparing the progress during the handoff (between t = 4s and t = 5s) with the progress following the handoff (between t=5s and t=6s). The results given in Table 1 summarize the effects of the handoff on TCP performance.

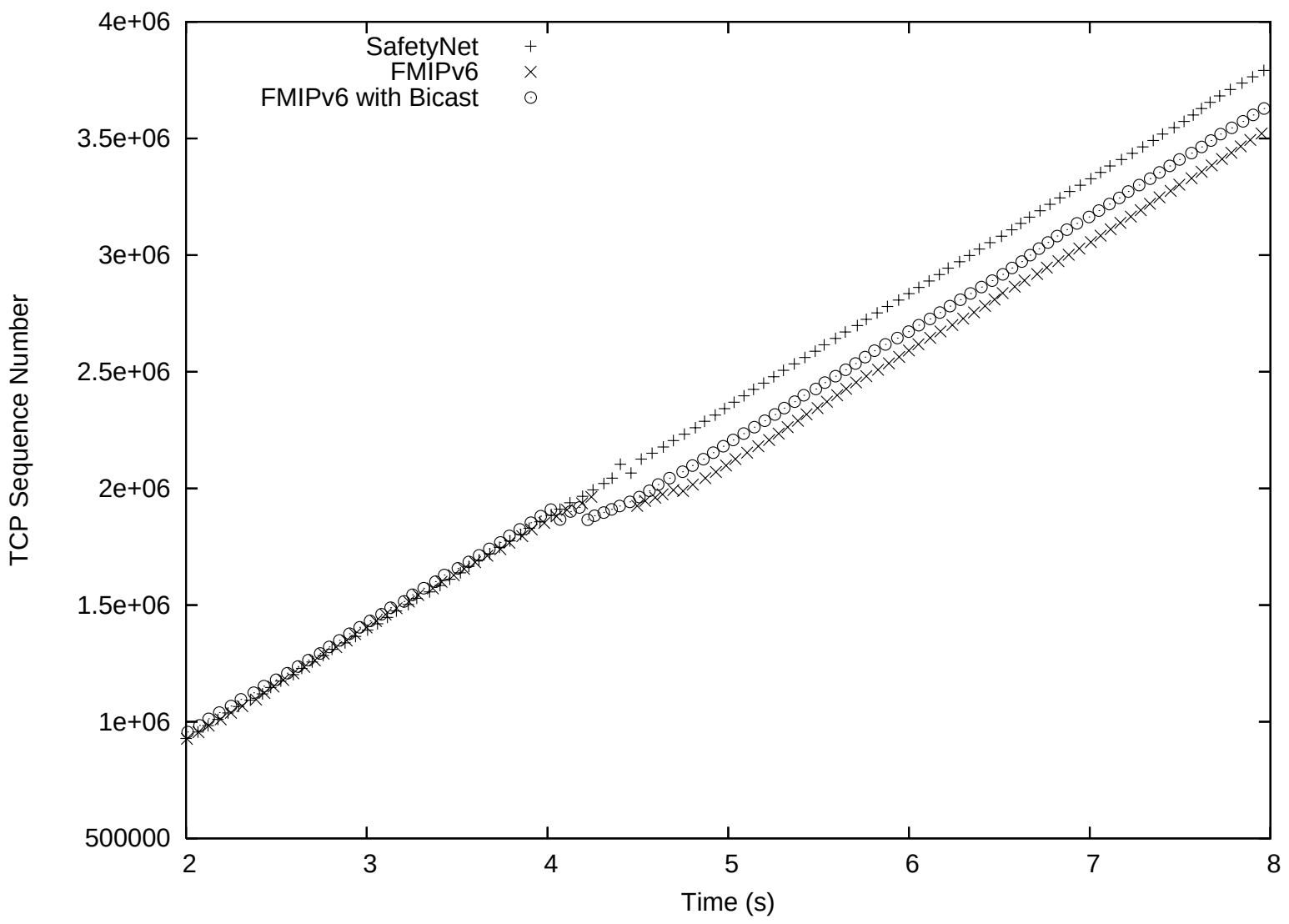

**Figure 5. TCP handoff performance comparison.**

**Table 1. Effect of handoff on TCP progress during a 1s window.**

| | *TCP progress* | *TCP resent data* | *Impact of handoff on TCP progress* |
|---|---|---|---|
| SafetyNet | 484154 Bytes | 2920 bytes | 0% |
| FMIPv6 | 246250 Bytes | 81100 bytes | 49% |
| FMIPv6 with Bicasting | 294650 Bytes | 86400 bytes | 39% |

More detailed sequence number graphs for SafetyNet and the FMIPv6 variants are given in Figure 6. The performance of FMIPv6 in Figure 6 a) illustrates the problems with buffering, when the speed of the data transfer is close to the capacity of the link. The first repeating of sequence numbers approximately at t=4.5 results from the sender (Correspondent Node) resending data after not receiving an acknowledgment in time due to the handoff delay. The slow emptying of the buffer due to the saturation of the link at the nAR causes TCP to start resending for a second time approximately at t=4.75s.

Figure 6 b) illustrates that FMIPv6 with bicasting suffers from a similar problem; although the Mobile Node can directly receive and acknowledge packets sent by the Correspondent Node during the handoff, duplicates of these packets are delivered to the buffer of the nAR. When the buffer is emptied and the duplicates are delivered at approximately t=4.1s, this saturates the link of the nAR and the delivery of fresh non-duplicate packets from the Correspondent Node is delayed. This delay causes the Correspondent Node to start resending packets due to a TCP time out and the resent packets are received at the Mobile Node at approximately t=4.2s. The receiving of duplicate packets also further degrade TCP performance, due to TCP interpreting the packet duplication as a sign of congestion. We saw this effect also in our earlier experimental results with the FMIPv6-BSD[9]. The improvements in SafetyNet over the FMIPv6-BSD protocol minimized the number of duplicate packets and removed this side effect of bicasting.

The performance of the TCP depicted in Figure 6 c) does not show any degradation from the SafetyNet handoff in spite of the resending of the two packets and the presence of a small number of duplicate packets. The amount of duplicate packets is minimized by the multi path delivery of the Stop Bicast message from the Mobile Node to the nAR, as described in the protocol description.

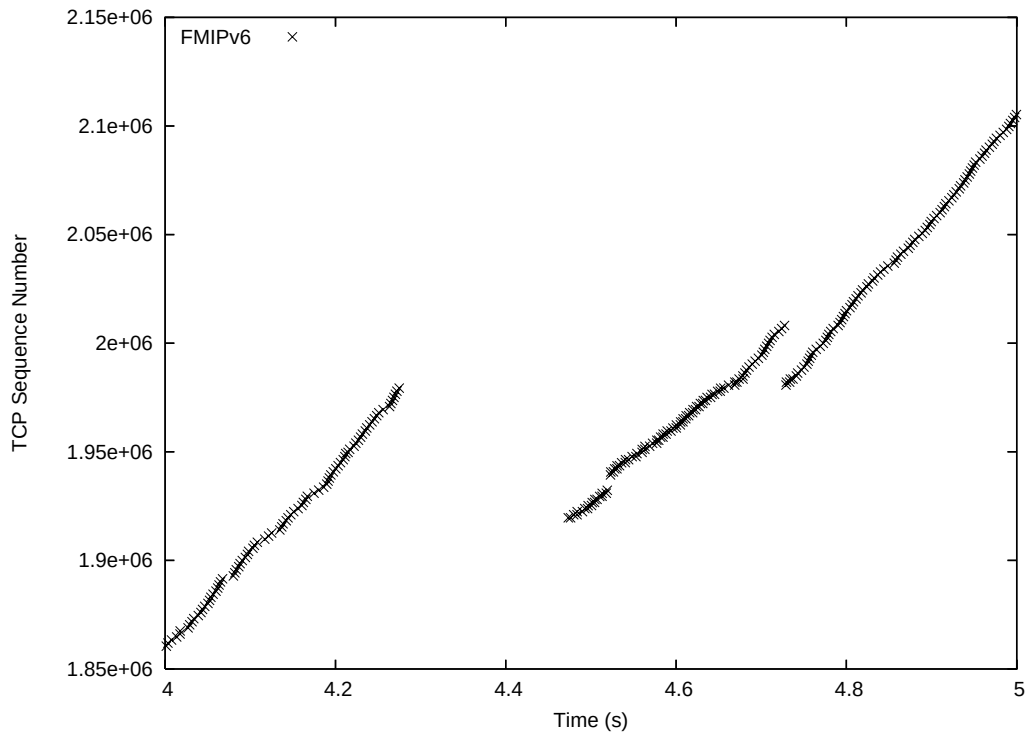

**a)**

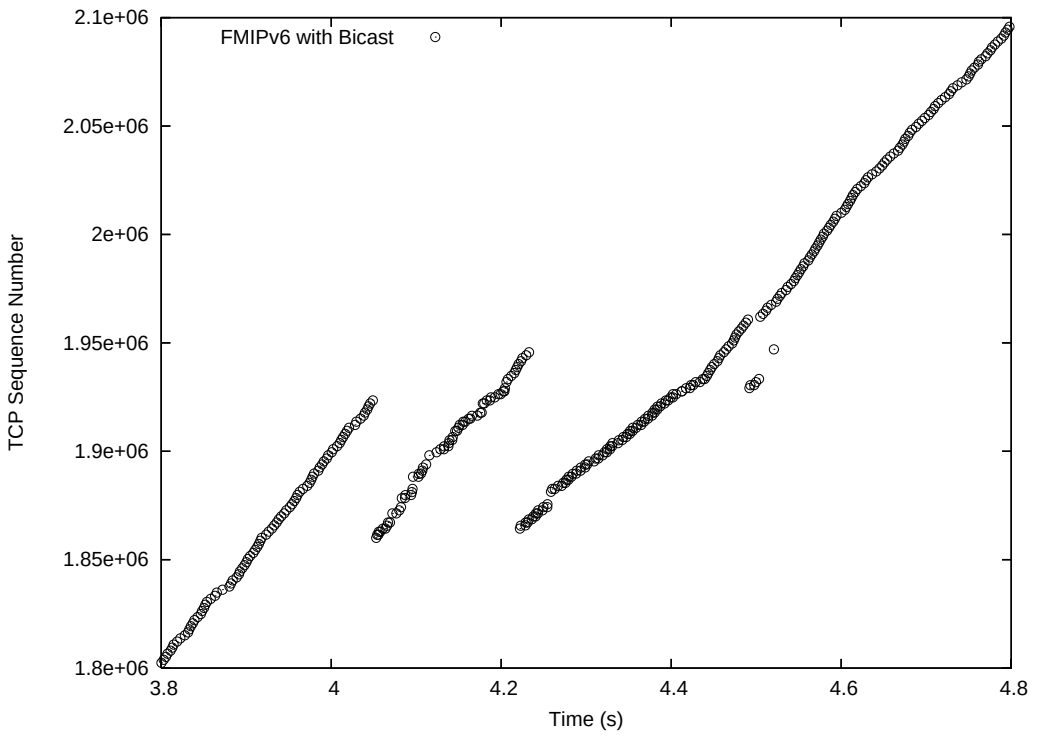

**b)**

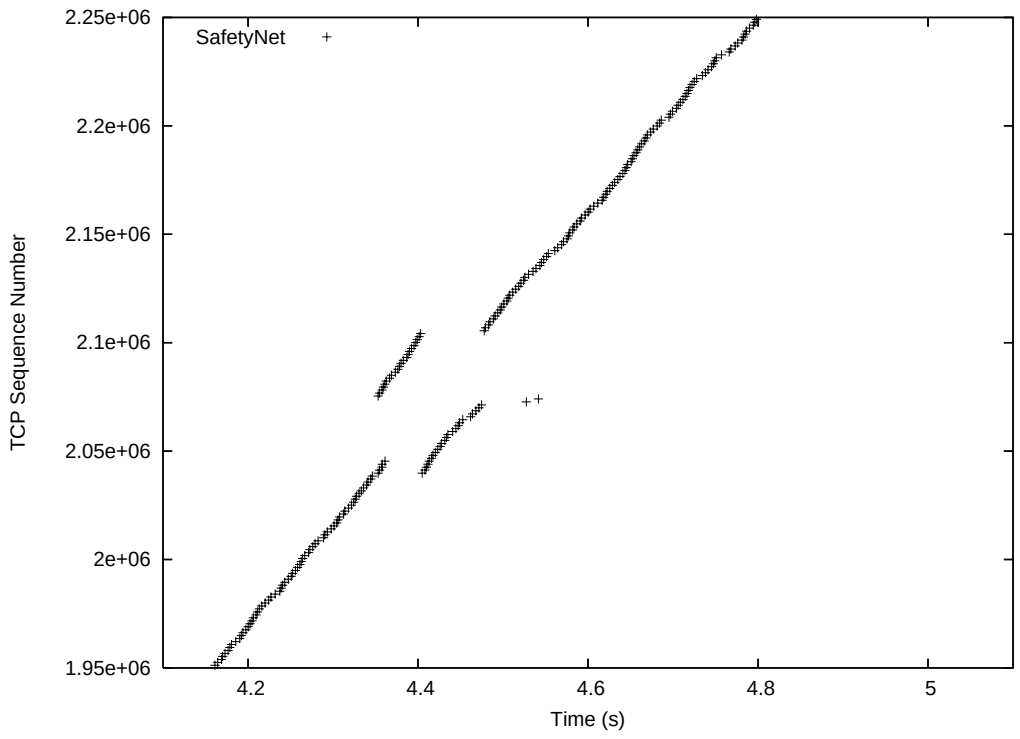

**c)**

**Figure 6. TCP handoff performance for a) FMIPv6, b) FMIPv6 with Bicasting and c) SafetyNet.**

The problems with link saturation are evident in the presented results in Figure 6 a) and b), even though in our test bed the old and new links use the same technology. In an upward vertical handoff, the decreased speed at the link of the nAR would lead to congestion and would thus affect the TCP flow regardless of the protocol used for the handoff. However, this would also amplify the negative impact of link saturation from buffering and thus the relative performance improvement of the SafetyNet protocol over the FMIPv6 variants would increase. However, in a scenario in which the link of the nAR was not the bottleneck, such as a downward vertical handoff, the performance of FMIPv6 would be closer to the performance to that of the SafetyNet protocol.

The results shown in this section did not show effects of packet loss at the link of the pAR during the handoff. With 100% packet loss during the handoff, SafetyNet protocol would perform similarly to FMIPv6, which would be the worst case performance for the SafetyNet protocol.

# 5 Numerical cost analysis of the SafetyNet protocol

The cost of SafetyNet consists of four main factors: The over-the air signaling and data transmission overheads and the over-the wire signaling and data transmission overheads. The over the air signaling overhead is the same as in FMIPv6 with the addition of the Stop Bicast message the Mobile Node sends directly to its pAR. The over the wire signaling overhead is the same as in FMIPv6 with the addition of the Stop Bicast message being sent from pAR to nAR. The size of the messages is given in Table 2 including the use of IPsec Authentication Header[13] for securing the messages.

**Table 2. Message Sizes for the SafetyNet and FMIPv6 protocols.**

| *Message Name* | *Message Size* | *With Authentication Header* |
|---|---|---|
| Fast Binding Update | 112 | 136 |
| Fast Binding Ack. | 72 | 96 |
| Handoff Initiate | 112 | 136 |
| Handoff Acknowledge | 72 | 96 |
| Router Sol. Proxy | 64 | 88 |
| Proxy Router Advert. | 80 | 104 |
| Fast Neighbor Advert. | 64 | 88 |
| Stop Bicast | 48 | 72 |

The over-the air signaling overhead is 584 bytes per handoff. If Mobile Node performs a handoff towards multiple routers, the handoff cost increases with 32 bytes per target router, due to the 2-way transfer of extra IPv6 addresses. The over the wire signaling overhead would be 304 bytes. The extra cost for multiple target nARs would be 232 bytes per target nAR, since the pAR and each nAR would exchange Handoff Initiate and Handoff Acknowledge messages.

The over-the air data transmission cost of the handoff in SafetyNet depends on the traffic used and the network topology. For evaluating vertical handoff performance, we divide the over the air cost to 1) over the air cost on the previous link and 2) over the air cost on the new link. The over the air data transmission cost on the previous link is the amount of data sent by the Correspondent Node during the handoff (between the sending of the Fast Binding Update and sending of the Fast Neighbor advertisement at the new link) and any packets delivered by the pAR between the end of the handoff and the pAR receiving the Stop Bicast message. The over the air data transmission cost over the new link due to the handoff is the amount of lost packets delivered from the buffer.

The over the wire transmission cost of the SafetyNet handoff between the pAR and the nAR is equal to the amount of packets sent by the Correspondent Node during the handoff and the overhead of tunneling, 40 bytes per packet. If the Mobile Node has multiple target nARs for the handoff, the over the wire cost is multiplied by the number of nARs. Each nAR needs buffer space for the estimated data delivered and potentially lost during the handoff.

We compare numerically the over the air overhead of our proposed SafetyNet protocol with FMIPv6 and FMIPv6 with bicasting. For the numerical evaluation, we use a TCP stream of 3.75 Mbps and a UDP stream of 100kbps as in our experimental evaluation. We analyze the total over the air cost of the protocols during a handoff using varying handoff latencies and a two different network latencies between the Mobile Node and the pAR for SafetyNet (t=5ms and t=50ms). The network latency of 5ms between the Mobile Node and the pAR corresponds to measured values from our test bed in the previous section and the latency of 50ms was chosen to represent the worst case scenario in a vertical handoff from a WLAN to a WWAN. In the worst case scenario, the directly sent Stop Bicast message from the Mobile Node would be lost and the pAR would only receive the message via the nAR. The costs of each protocol are compared against the data transmitted over the handoff period in Figure 8 for both UDP and TCP.

The total over the air costs of a handoff for each protocol when taking TCP resending into account are given in Figure 9 for our experimental scenario. The handoff cost is derived from Figure 8 b) by setting the handoff latency to 200ms and the network latency between the Mobile Node and the pAR to 5ms. We used the experimental data acquired in Section 4.2 to derive the TCP resend cost. The resend cost does not include the cost of duplicate messages which are already taken into account in the handoff cost along with the tunneling overhead and the signaling overhead.

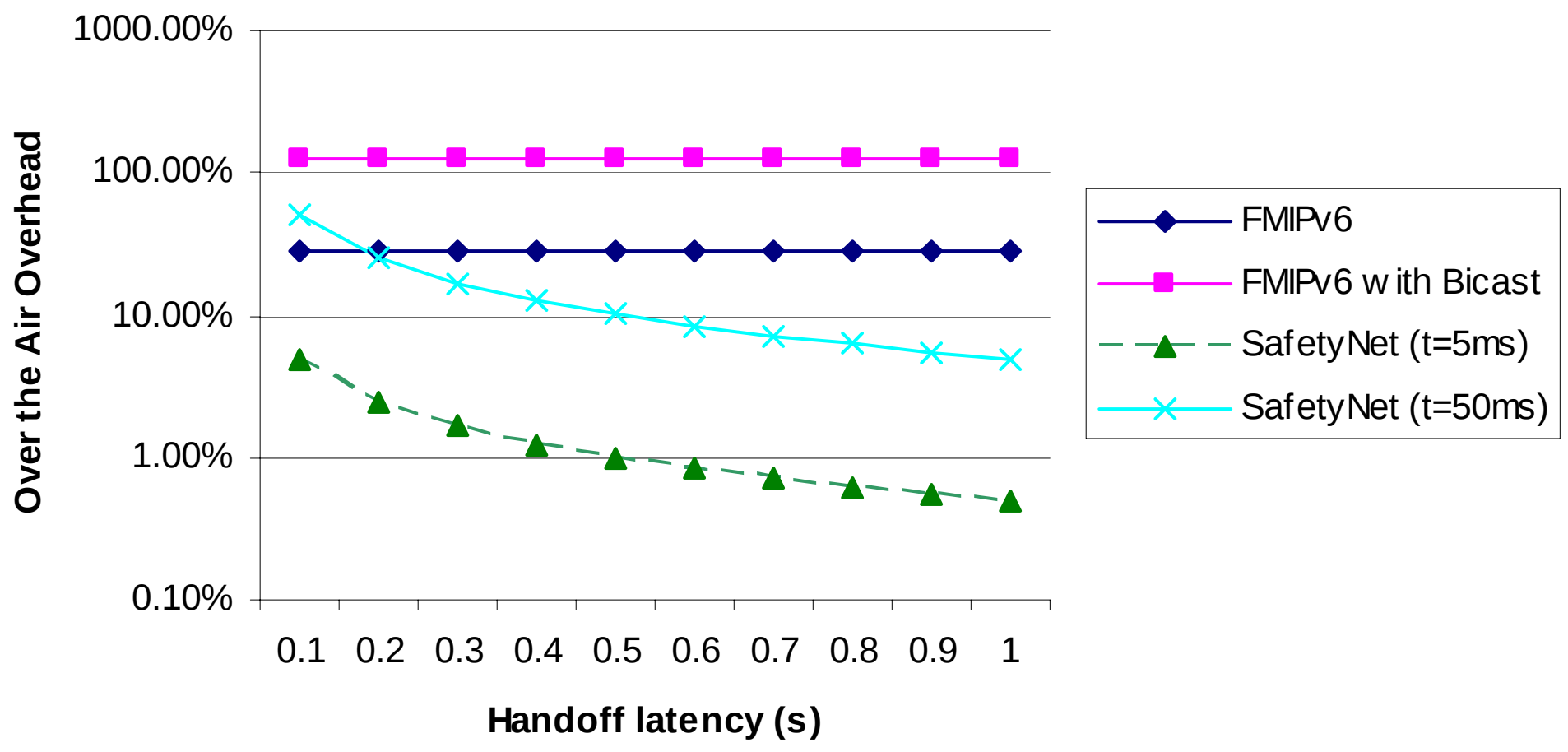

**a) UDP Handoff Overhead**

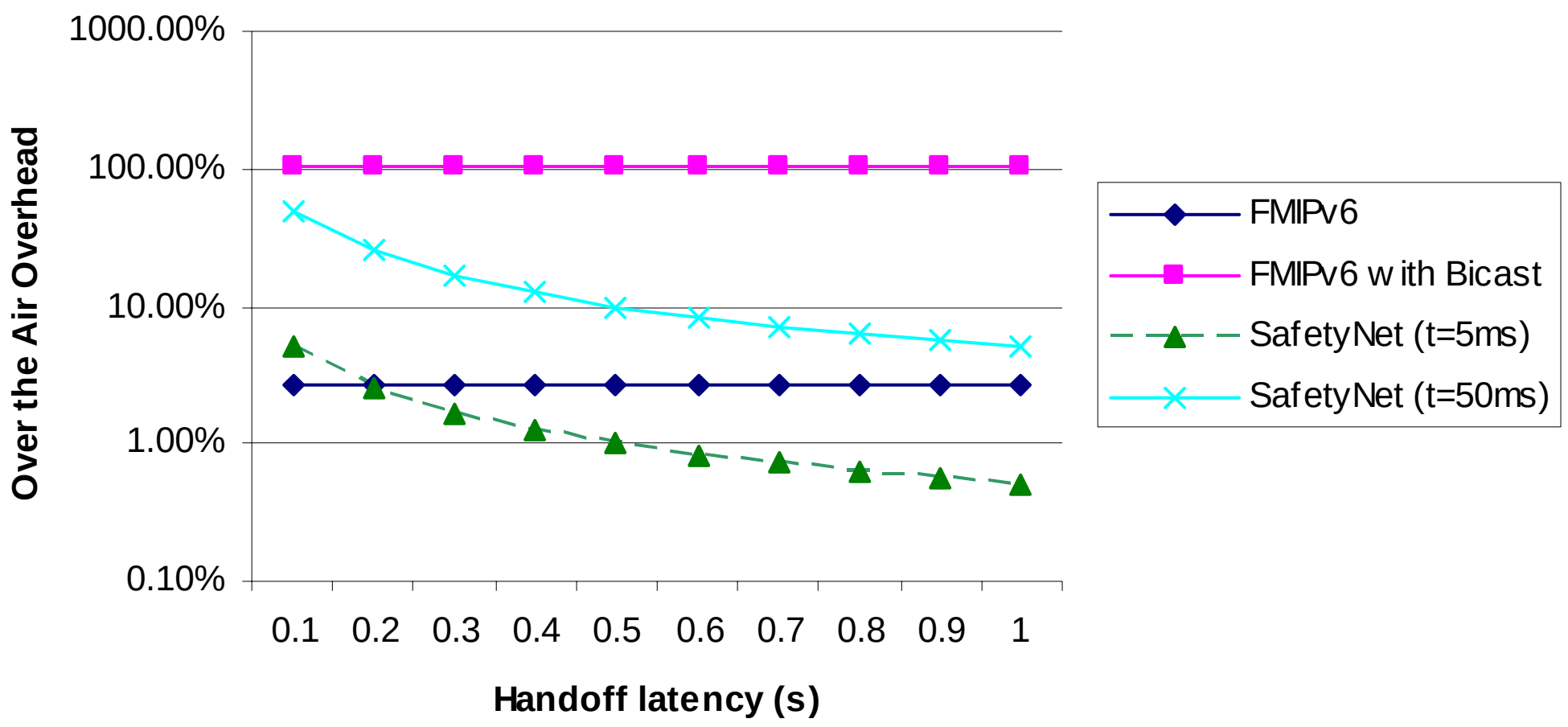

**b) TCP Handoff Overhead**

**Figure 7 Comparison of the over the air signaling and data transmission overhead for a) UDP and b) TCP.**

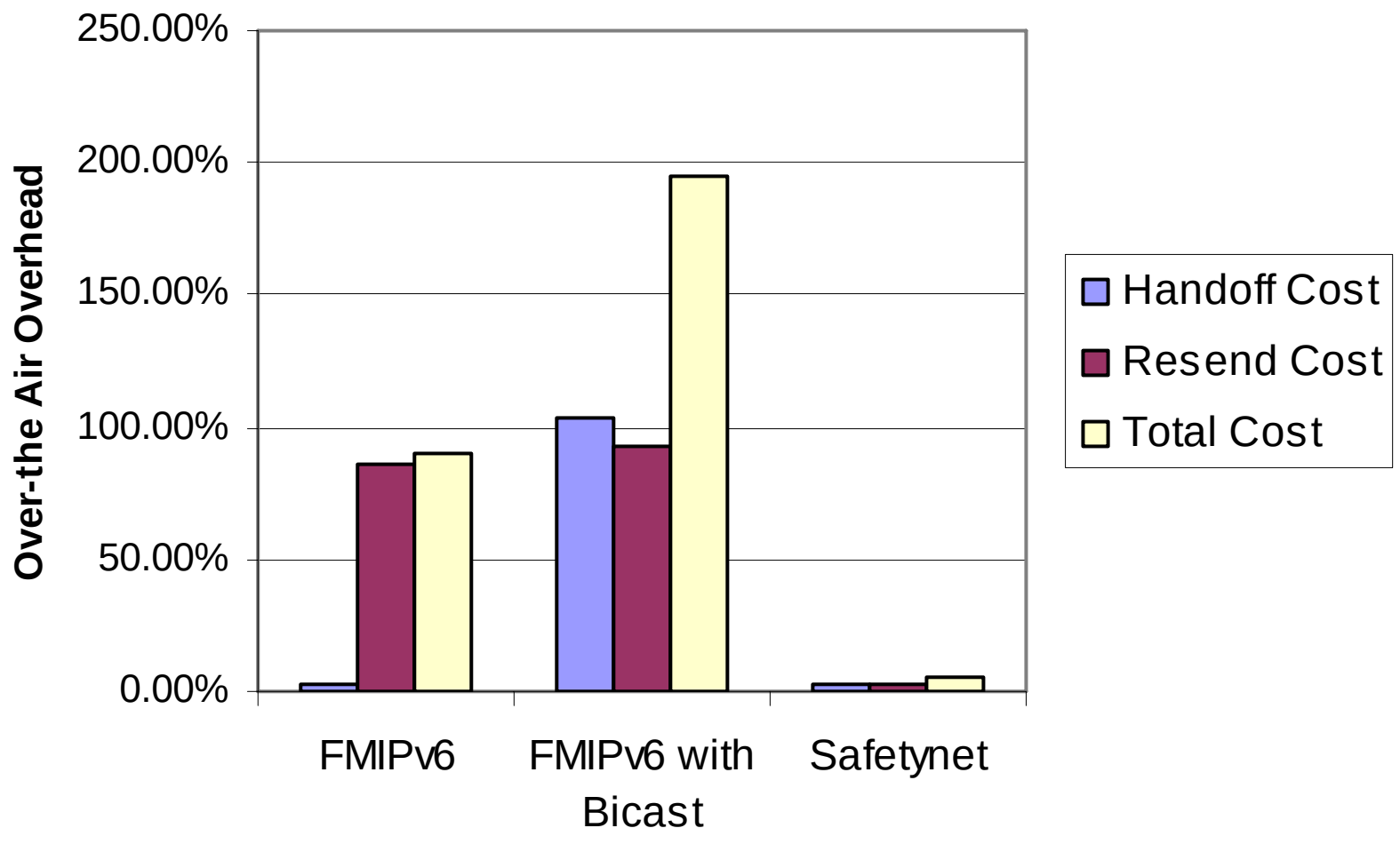

**Figure 8. Experimental over the air data transmission, signaling and TCP resend overhead comparison.**

# 6 Conclusions

In this paper, we presented the SafetyNet protocol which enables seamless vertical handoffs. SafetyNet provides a localized mobility management scheme for vertical handoffs allowing a Mobile Node to receive any packets lost during the handoff at the new Access Router. Further, selective delivery from the buffer at the new Access Router guarantees that only the lost packets are sent to the Mobile Node reducing the overhead of the handoff significantly. The SafetyNet protocol together with the proposed handoff timing algorithm would allow a Mobile Node to delay the finalization of an upward vertical handoff to a WWAN network and in some cases, to even avoid performing such a handoff. This would enable a Mobile Node to increase its utilization of low cost WLAN networks.

The over the air costs of a protocol can be considered as one of the key metrics of a wireless protocol. The SafetyNet protocol is designed in a manner in which the total over-the costs are minimized. The total over the air overhead caused by SafetyNet handoff that we saw in our experiments was 93% smaller than the overhead from FMIPv6 and 97% smaller than the overhead of FMIPv6 with bicasting. This, together with the improvement in the handoff performance of up to 95% for TCP performance in vertical handoffs, when compared with FMIPv6 and an improvement of 64% over FMIPv6 with bicasting confirms the viability of the proposed SafetyNet protocol.

## Acknowledgments

The authors would like to thank Max Ott for his helpful comments.